\documentclass[twocolumn,showpacs,preprintnumbers,amsmath,amssymb]{revtex4}
\usepackage{tabularx,graphicx}\begin{document}
\newcommand{\bk}{\bold{k}}
\newcommand{\bkp}{\bold{k'}}
\newcommand{\beq}{\begin{equation}}
\newcommand{\eeq}{\end{equation}}
\newcommand{\beqn}{\begin{eqnarray}}
\newcommand{\eeqn}{\end{eqnarray}}
\newcommand{\bmath}{\begin{subequations}}
\newcommand{\emath}{\end{subequations}}
\title{Materials and mechanisms of hole superconductivity}
\author{J. E. Hirsch }
\address{Department of Physics, University of California, San Diego,
La Jolla, CA 92093-0319}

\begin{abstract} 
 The theory of hole superconductivity proposes that there is a single mechanism of superconductivity that applies to all superconducting materials. This paper discusses several material families where superconductivity occurs and how they can be understood within this theory. Materials  discussed include the elements, transition metal alloys, high $T_c$ cuprates both hole-doped and electron-doped, $MgB_2$, iron pnictides and iron chalcogenides, doped semiconductors, and elements under high pressure.   \end{abstract}

\maketitle 

The conference series ``Materials and Mechanisms of Superconductivity'' started in 1988, at the dawn of the high $T_c$ cuprate era, and has
given rise to 
9 international meetings so far. As its  name implies, it assumes that more than one  mechanism of superconductivity is required to explain the large 
variety of superconducting materials found so far.   Instead, we have proposed\cite{hole} that there is a single mechanism to explain superconductivity in $all$ materials, both materials already discovered as well as those to be discovered, 
that  is $not$ the electron-phonon interaction. None of the other proposed new mechanisms of superconductivity
questions the validity of the conventional BCS-electron-phonon mechanism for conventional superconductors.

Our theory, ``hole superconductivity'', proposes that superconductivity is only possible when hole carriers exist in the metal\cite{hole1}, that 
superconductivity results from Coulomb rather than
electron-phonon interactions\cite{kl}, and that it  is particularly favored (yielding high  $T_c$ 's)   when  holes conduct through a 
network of closely spaced {\it negatively charged anions}, as in the $Cu^{++}(O^=)_2$ planes shown in Fig. 1 \cite{tang,hsc1}. Here  I discuss the superconductivity of several
classes of materials in the light of these principles.

   \begin{figure}
 \resizebox{6.0cm}{!}{\includegraphics[width=9cm]{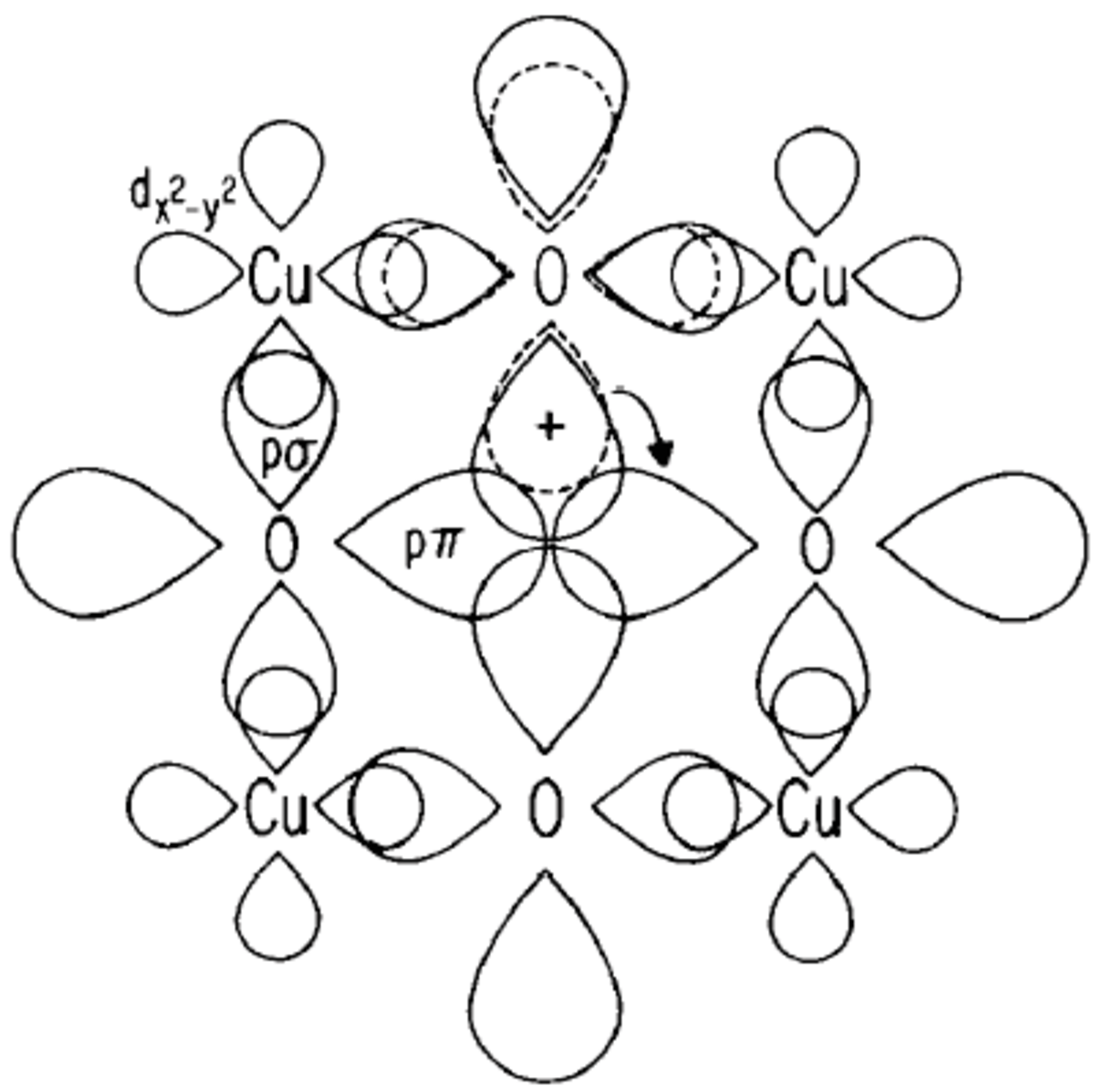}}
 \caption {Carriers responsible for high $T_c$ superconductivity in the cuprates reside in 
 oxygen $p\pi$ orbitals in the $CuO$ planes according to the theory of hole superconductivity\cite{tang,hsc1}.   }
 \label{figure2}
 \end{figure}

\section {`conventional' superconductors}
We denote by `conventional' superconductors those that are generally believed to be described by conventional 
BCS-Eliashberg theory. 

Almost all superconducting $elements$ have positive Hall coefficient in the normal state, indicating that hole carriers dominate
the transport. Examples are {\it Pb, Al, Sn, Nb, V, Hg}. Instead, most elements with negative Hall coefficient are
non-superconductors at ambient pressure down to the lowest temperatures checked so far,
for example {\it Ca, Sc, K, Mg, Ag, Au}. This was noted by several workers in the early days of superconductivity\cite{hall}. 
The $sign$ of the Hall coefficient is the strongest normal-state indicator of superconductivity among thirteen normal state
properties considered in ref. \cite{correl}. This is not accounted for by BCS theory, and normal state properties expected to be
related to superconductivity within BCS theory (like ionic mass, Debye temperature, specific heat and resistivity) show substantially
weaker correlation with superconductivity\cite{correl}.

The behavior of the transition temperature of alloys of transition metals in different columns of the periodic table versus 
composition shows characteristic behavior that  can be understood by a `universal' curve in terms of electrons per 
atom ratio ($e/a$). This is known as `Matthias' rules'\cite{matthias}, and is shown in the top panel of Fig. 2 for a large number of transition metal
alloys with $e/a$ ratio between $4$ and $6$ (from ref. \cite{tcvse/a}).  It can be simply understood from the carrier density dependence of $T_c$ in
a simple one-band model for hole superconductivity\cite{transition1}. The pairing interaction is given by 
\beq
V_{\bk\bkp}=U+V(\bk-\bkp)-\alpha (\epsilon_\bk+\epsilon_{\bkp})
\eeq
where $\epsilon_\bk$ is the band energy measured from the center of the band,  $U$ and $V$ are on-site and
more extended Coulomb repulsions and $\alpha>0$ arises from `correlated hopping'\cite{correlhop},
an electron-electron interaction term that is proportional to the hopping amplitude\cite{isotope}. The interaction Eq. (1) becomes progressively  less repulsive as the Fermi level goes up in the band
(as $\epsilon_\bk$, $\epsilon_\bkp$ for $\bk$ and $\bkp$ at or near the Fermi surface increase).
Superconductivity
arises as the Fermi level approaches the top of the band, $T_c$ increases as the pairing interaction gets stronger with
increasing band filling, reaches a maximum, starts decreasing as the number of carriers (holes) becomes small, and
reaches zero when the Fermi level crosses the top of the band. This is indicated in the anomaly in the Hall coefficient shown
in the bottom panel of Figure 2 (from ref. \cite{hallvse/a}). Figure 3 shows the calculated band structure for elements
$Ti, V$ and $Cr$, corresponding to $e/a=4, 5$ and $6$ respectively
(from Ref. \cite{moruzzi}). $T_c$ goes to zero as the Fermi level crosses the
top of the band at the $\Gamma$ point and the hole pocket disappears. A    calculation  using a realistic
band structure   and interactions of the form Eq. (1) reproduces this behavior closely\cite{transition1}.

   \begin{figure}
 \resizebox{7.5cm}{!}{\includegraphics[width=7cm]{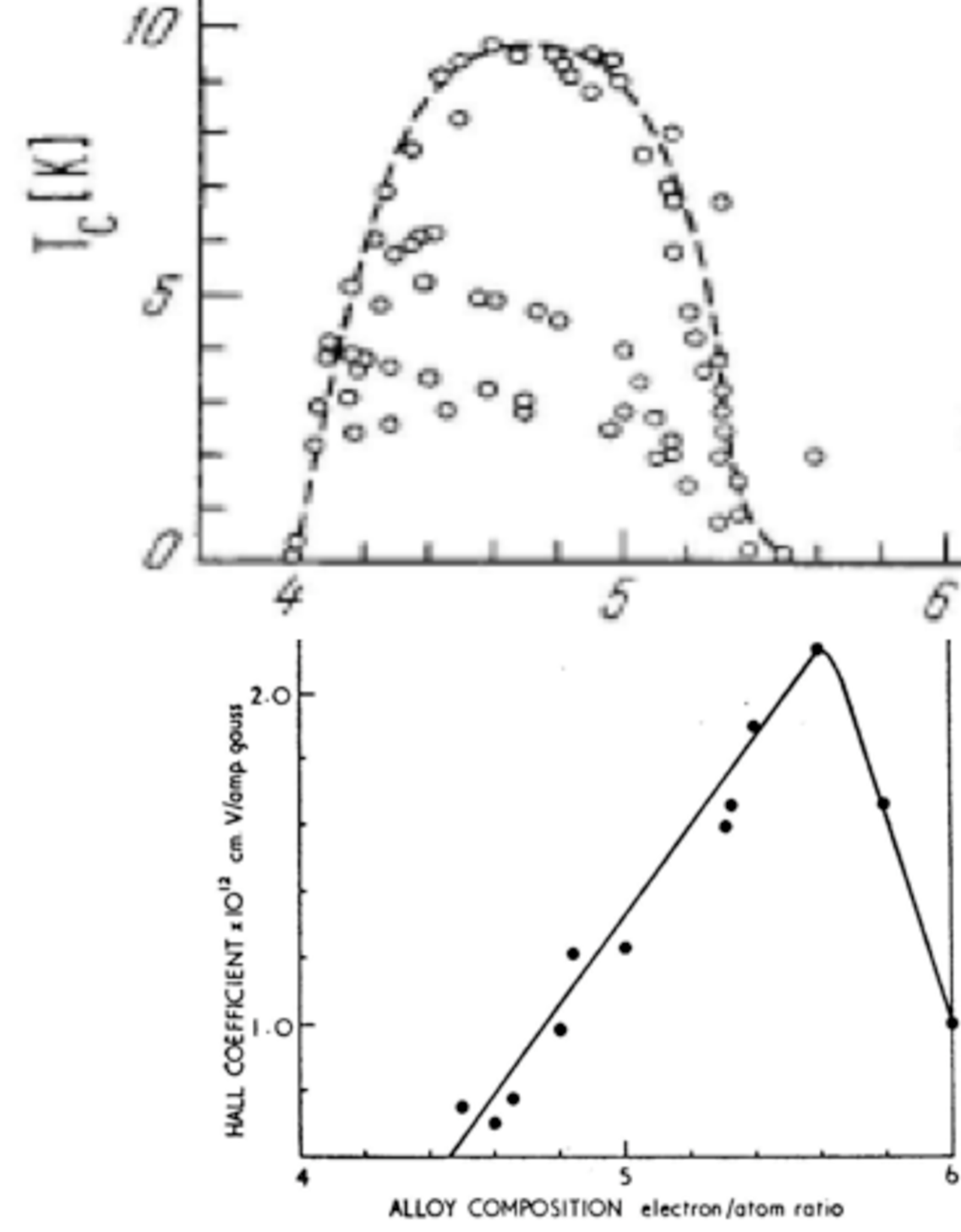}}
 \caption {$T_c$ versus e/a ratio (from ref. \cite{tcvse/a} (top panel)) and Hall coefficient versus e/a ratio (from ref.\cite{hallvse/a})
 (bottom panel. Note that $T_c$ goes to zero at the point where the Hall coefficient shows a pronounced kink, for $e/a\sim 5.6$.   }
 \label{figure2}
 \end{figure}
 
    \begin{figure}
 \resizebox{7.5cm}{!}{\includegraphics[width=7cm]{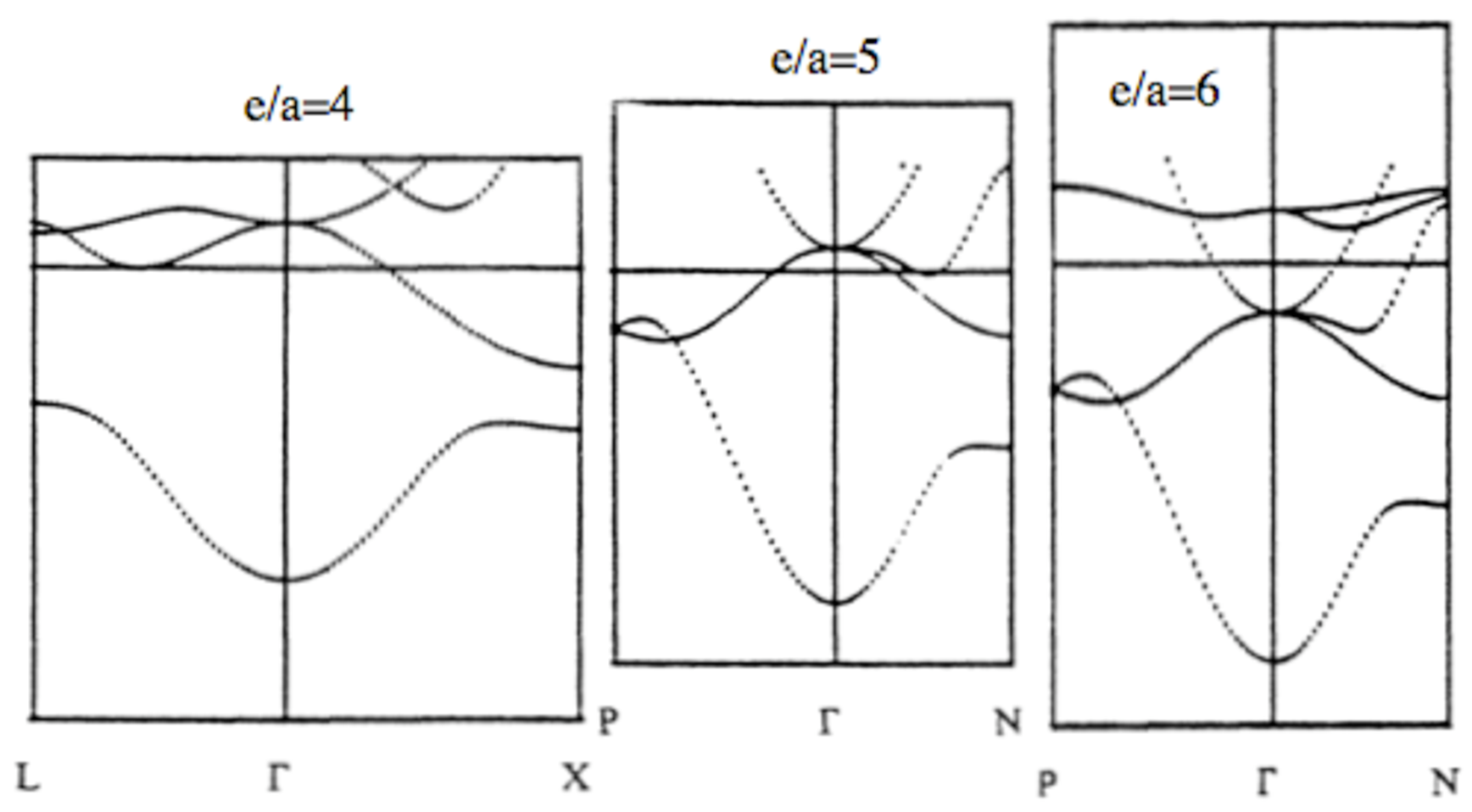}}
 \caption { Band structures of elements $Ti, V, Cr$ with $e/a=4, 5, 6$ in the fourth row
 of the periodic table, from ref. \cite{moruzzi}.   The horizontal line denotes the position of  the Fermi level.}
 \label{figure2}
 \end{figure}

Magnesium diboride ($MgB_2$) is a textbook example of the mechanism of hole superconductivity at work\cite{holemgb2}. 
The system consists of  negative ions $B^{-}$ forming planes, separated by arrays of
positive $Mg^{++}$ ions. The charge transfer is however not complete, and as a result a small density of hole carriers exist in
the $B^{-}$ planes, some of which reside in planar $p_{xy}$ orbitals and propagate through direct overlap of
these orbitals shown schematically in the left panel of Fig. 4. The right panel of Fig. 4 shows the resulting
 small hole pocket at the $\Gamma$ point that 
gives rise to a cylindrical Fermi surface describing hole conduction in the $B^{-}$ 
planes\cite{mgb2bands}. 
There is also electron conduction in this system in three-dimensional bands involving boron $p_z$ orbitals
and $Mg^{++}$ orbitals.

Tunneling experiments show the existence of two superconducting gaps, as shown in the left panel of Fig. 5 \cite{mgb2tunn}.
The larger gap is associated with hole carriers propagating through the $B^{-}$ planes, and the smaller gap
is associated with a three-dimensional band with electron carriers. Several years earlier Marsiglio and
the author  calculated the superconducting properties of a two-band model with holes in one band and
electrons in the other band within the model of hole superconductivity\cite{twoband}, and  found a behavior for the gaps
similar to the one seen in $MgB_2$, as shown on the right panel of Fig. 5. 

   \begin{figure}
 \resizebox{9.0cm}{!}{\includegraphics[width=9cm]{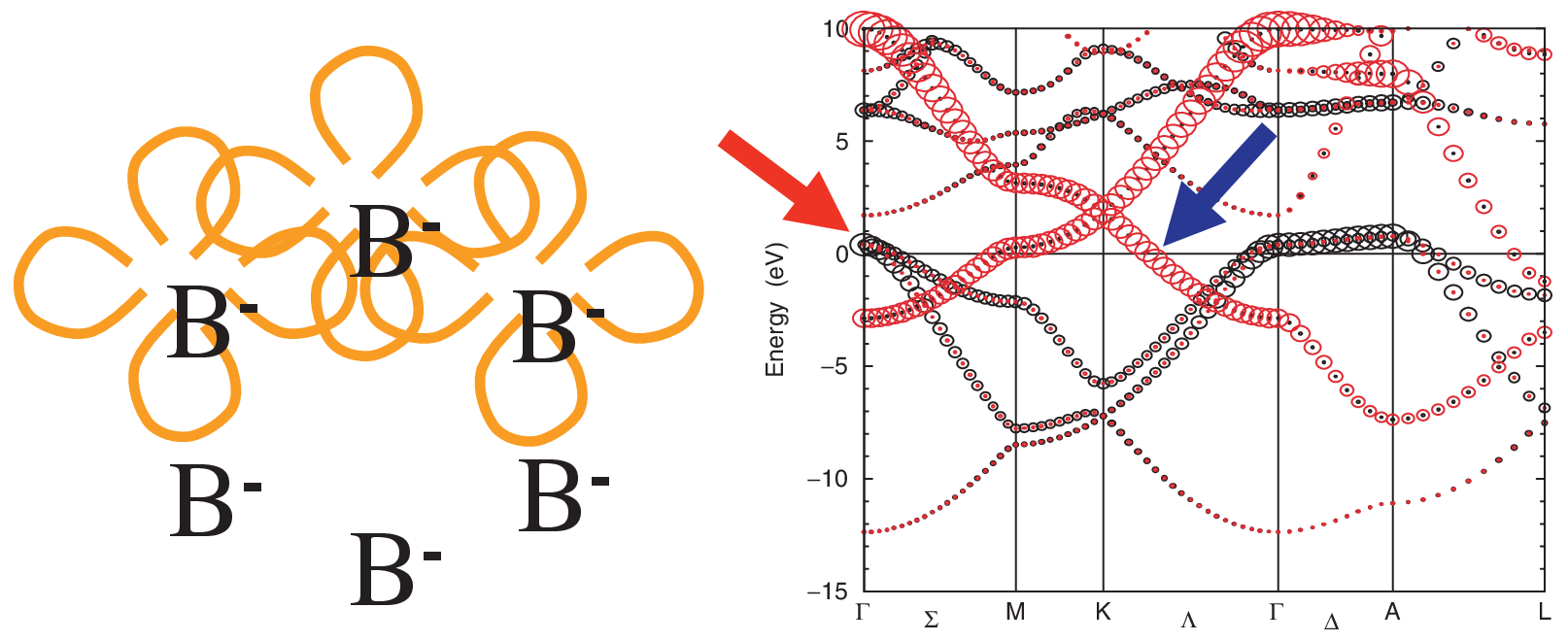}}
 \caption {The left panel shows schematically the boron $p_{xy}$ orbitals in the $B^-$ planes, where conduction occurs through
 holes  in the pocket near the $\Gamma$ point indicated by the red arrow on the right panel, that shows the band structure
 of $MgB_2$ from Ref. \cite{mgb2bands}. The blue arrow shows a three-dimensional band where the carriers are electron-like.  }
 \label{figure2}
 \end{figure}
 
   \begin{figure}
 \resizebox{9.0cm}{!}{\includegraphics[width=9cm]{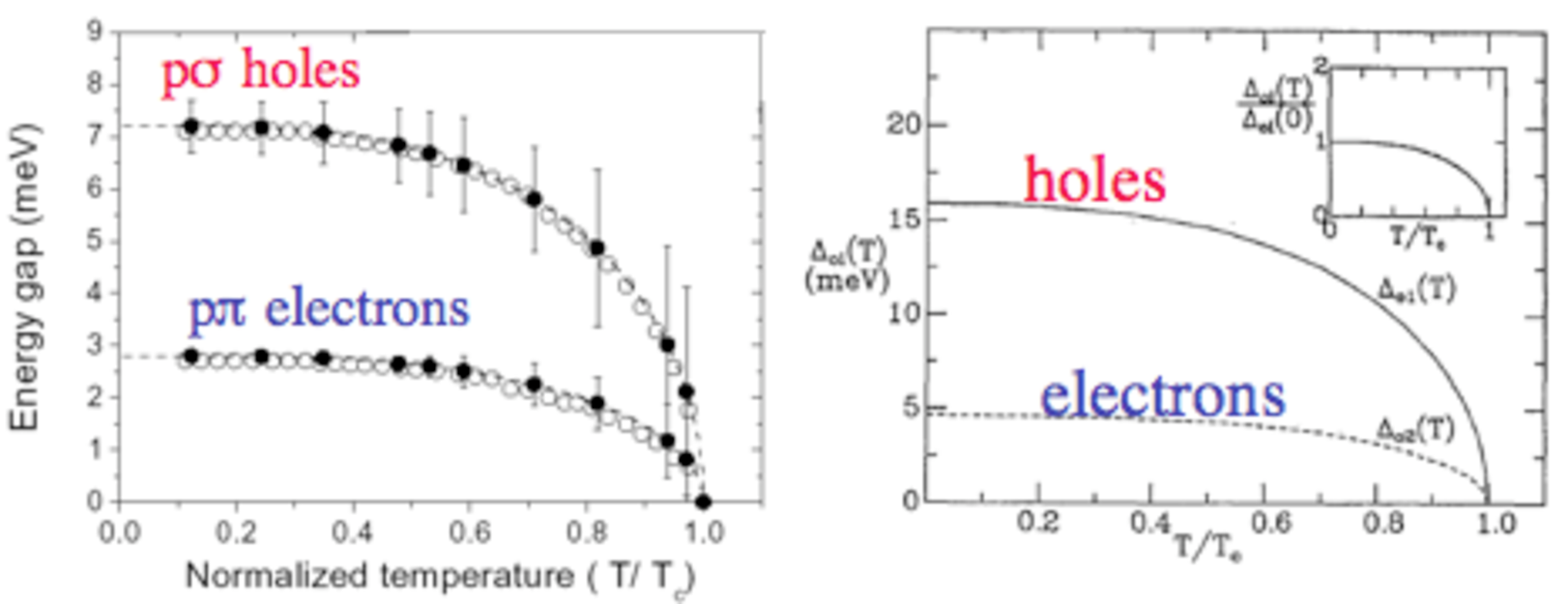}}
 \caption {Left panel: two gaps versus temperature for $MgB_2$ obtained from NIS tunneling\cite{mgb2tunn}.
 The large and small gap correspond to the band structure states indicated by the red and blue arrows respectively in
 the left panel of Fig. 4. The right panel of this figure shows results of a two-band model calculation within the theory of
 hole superconductivity obtained 10 years before the discovery of $MgB_2$\cite{twoband}.   }
 \label{figure2}
 \end{figure}

In the model of hole superconductivity, the transition temperature is higher when hole conduction occurs through
{\it negatively charged ions}\cite{hole1,hole2}. Thus, the fact that $T_c$ is so high in $MgB_2$ compared to other $s-p$ 
superconductors derives from this feature together with the hole conduction.

\section {`unconventional' superconductors}
By `unconventional' superconductors we denote those that are generally believed to be $not$  described by conventional 
BCS-Eliashberg theory. The 
 high $T_c$ cuprates (both hole- and electron-doped), iron pnictides and
   iron selenium/tellurium show features clearly consistent with the
mechanism of hole superconductivity. Other superconductors 
generally agreed to be `unconventional' are heavy fermion materials and stronthium ruthenate. Because their $T_c$ is
so low, it is difficult to find clear signatures in favor of our mechanism for those materials.

We will skip here  a  
discussion of the cuprates,  which we have discussed in detail in several
papers\cite{tang,hsc1,correlhop,marsiglio,london,stm,houston,photo,hsc2,twoband}.  Let us just mention that our model $predicted$\cite{edoped} that in electron-doped cuprates hole
 carriers exist and are responsible for superconductivity  well $before$  this
 prediction was supported by detailed transport measurements\cite{greene}.
 
 For iron pnictides, we have discussed the mechanism by which hole carriers are expected
 to be generated both through electron- or hole-doping\cite{feas}. For electron doping, the mechanism
 is similar to the one proposed for the electron-doped cuprates\cite{edoped}.
 The   negatively charged anion $As^{---}$ is the key element in these superconductors to
 give rise to the high transition temperatures.

The
iron-selenium system ($FeSe$) is another textbook example of the applicability of our model\cite{fese}. Under pressure, this system
increases its critical temperature from $8K$ to $37K$ for pressures in the range 6-9 GPa\cite{fese3,fese4}. The main effect of pressure is to decrease the
distance between $Se$ atoms in neighboring planes as shown schematically in Fig. 6, from $3.68 \AA$ to $3.16\AA$\cite{fese,fese2}. 
The Pauling radius of the $Se^=$ ion is $1.98 \AA$, so the ions increase their overlap substantially under pressure,
which should lead to a large increase 
in $T_c$ according to the model of hole superconductivity. Indeed, in discussing
the  large positive pressure dependence of $T_c$ in the cuprates over 20 years ago we proposed that it arises from increase of
the overlap of the $O^=$ ions in the $Cu-O$ planes according to the model of hole superconductivity\cite{marsiglio}.

Note also that if the $Fe$ ion would play a substantial role in the superconductivity, as expected within other theories, it would
be difficult to explain the very large effect of pressure in increasing the critical temperature since the change in distance
within a $Fe-Se$ plane (e.g. the $Fe-Fe$ distance decreases  from $2.67\AA$ to $2.60\AA$,  $\sim 2\%$) is much smaller than 
the distance decrease between $Se$ atoms in adjacent planes ($3.69\AA$ to $3.17\AA$,  $\sim 16\%$).

   \begin{figure}
 \resizebox{8.5cm}{!}{\includegraphics[width=9cm]{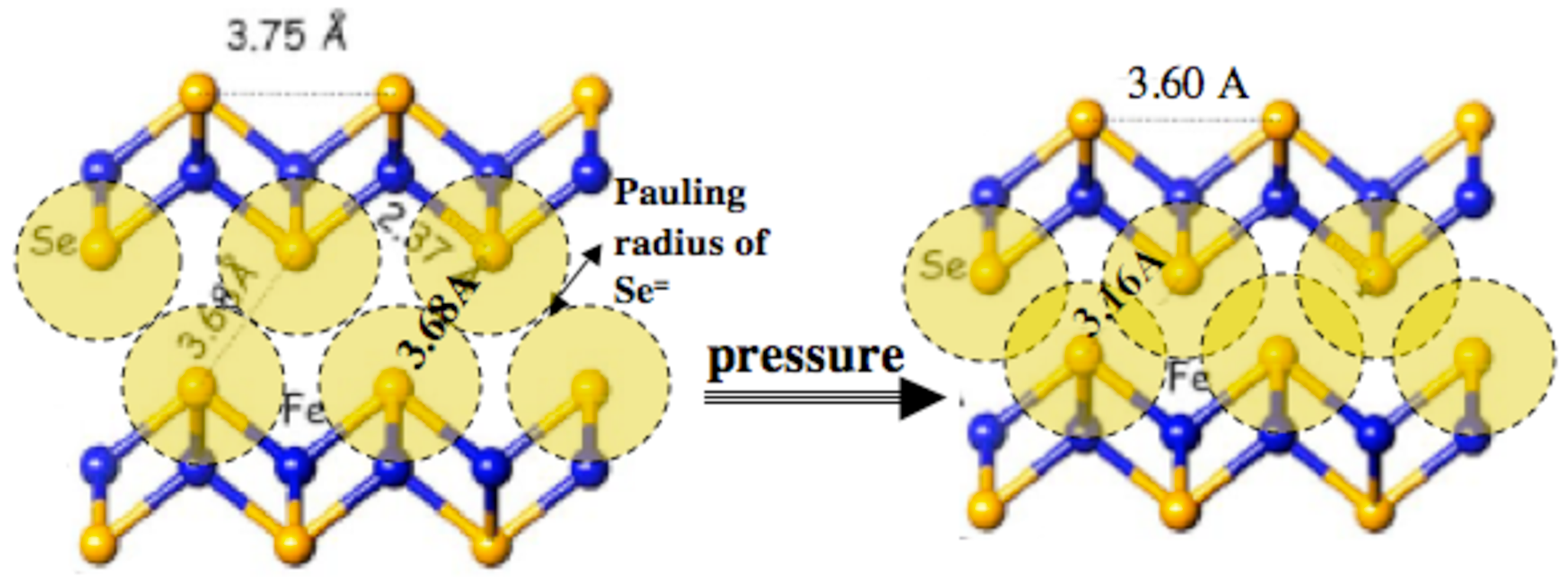}}
 \caption {Schematic depiction of $FeSe$ planes without (left) and with (right) application of pressure (part of this figure was reproduced from Ref. \cite{fese2}, Fig. 4(d)). 
 The main effect of pressure
 is to reduce the $Se^=-Se^=$ distance between $Se^=$ anions in neighboring planes, leading to substantial overlap of 
 anion orbitals.}
 \label{figure2}
 \end{figure}
 
    \begin{figure}
 \resizebox{8.5cm}{!}{\includegraphics[width=9cm]{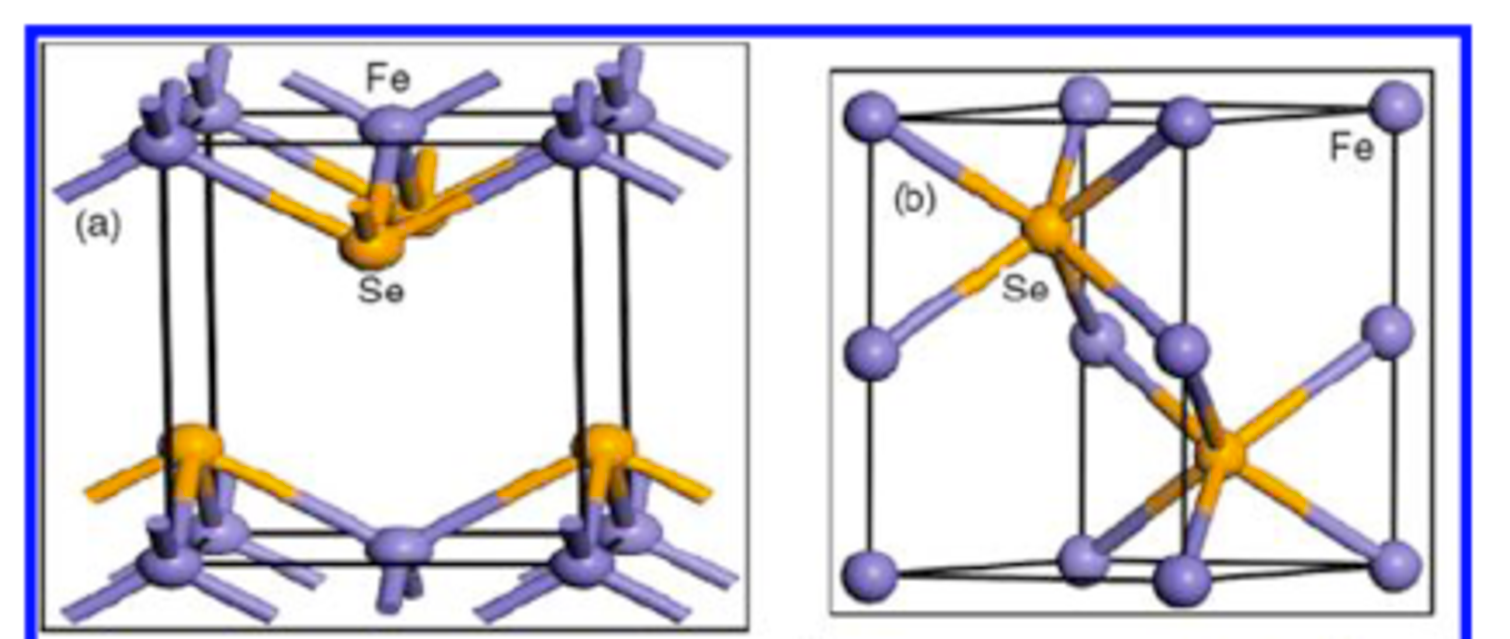}}
 \caption {Low pressure (left) and high pressure (right) phases of $FeSe$
 (from ref. \cite{fese}. Unlike the low pressure phase,
 in the high pressure phase conduction occurs always through $Fe$ sites since there is no direct overlap of $Se$ orbitals.}
 \label{figure2}
 \end{figure}
 
At higher pressures, the system undergoes a structural transition and is no longer superconducting.
 Figure 7 shows the structures before and after the transition. It can be seen that the high pressure phase (right panel in Fig. 7) does not
 allow for direct overlap of anion orbitals, therefore it is expected within the theory of hole superconductivity that
 it will not be a high $T_c$ superconductor.
 
 Finally, note that the compound $SnO$, with no traces of magnetism, has the same structure and similar band structure as
 $FeSe$ and $is$ a superconductor\cite{sno}. This is consistent with our model and inconsistent with theories that assume magnetic
 fluctuations play an important  role in the superconductivity of these materials\cite{scal}.

\section {`undetermined' superconductors}
For several classes of materials there is no consensus in the community whether they are `conventional' or
`unconventional'. Among the members of this class we will discuss the recently discovered hole-doped
semiconductors   and simple metals under high pressure.

It has been found in recent years that doping diamond, Si and Ge with $holes$ gives rise
to superconductivity\cite{semicond}. This is consistent with the theory of hole superconductivity, as is the
fact that superconductivity is $not $ found when   these semiconductors are doped with $electrons$.
BCS-Eliashberg theory did not predict the existence of superconductivity in these materials
upon hole doping, nor does it explain why electron doping does not give rise to
superconductivity. The $T_c$ is quite low (maximum is $11.4K$)  consistent with the
prediction of the theory that requires negatively charged ions in addition to hole carriers
to give rise to high temperature superconductivity, as well as with the fact that
the coordination number in these materials is quite small which disfavors a high $T_c$ within our model\cite{correlhop}.

The rather high $T_c's$ recently found in simple and early transition metals under high pressure are claimed
to be explained by conventional theory but were not predicted by it. Examples are 
Li ($T_c=20K$), Ca ($T_c=25K$), Sc ($T_c=19.6K$) and Y ($T_c=19.5K$). Within our theory
superconductivity occurs in these materials\cite{hamlin} because under application of pressure new
Bragg planes develop that convert electron carriers into hole carriers\cite{deg}. We predict that the Hall coefficient
(not yet measured) of these materials under pressure will  change sign from negative to positive in the range of pressures
where they become superconducting\cite{hamlin}, or at least that there will be  clear evidence in magnetotransport studies for two-band conduction, with one of the
carrier types being hole-like. If this is not observed it would cast serious doubt on the validity of the theory.

\section {discussion}
The theory of hole superconductivity is qualitatively different from BCS-Eliashberg theory. 
Thus for many materials the predictions of both theories will disagree. 
For example, 
contrary to BCS-Eliashberg theory\cite{pickett} we predict no
high $T_c$ superconductivity in hole-doped  LiBC because there is no conduction through 
overlapping orbitals of negatively charged ions.
We also predict no superconductivity in metallic hydrogen $unless$ the structure distorts to accommodate an even number of atoms per unit cell,
in contrast to conventional BCS  theory that predicts high $T_c$ with no lattice distortion\cite{ashcroft}.
However for some materials both theories  could agree on their predictions.
For example, BCS theory predicts that superconductivity is favored when there is a 
soft phonon mode\cite{soft}. A soft phonon mode is often a precursor to a lattice instability 
and is likely to occur when there are   $antibonding$ electrons that disfavor the
stability of the solid because they give rise to low charge density in the region between the
atoms. At the same time, antibonding electrons reside in the high states in the band,
hence give rise to hole carriers. We argue that  superconductivity appears to
be favored by soft phonon modes because it is the antibonding electrons that are responsible for  $both$ the
(hole) superconductivity $and$ the existence of soft phonon modes.

 Within the theory of hole superconductivity all superconductors should be explainable by the
 same mechanism. A single superconductor that demonstrably does not fit the requirements of
 the theory, for example a superconductor that does not have any hole-like carriers in the normal state, 
  would prove the theory wrong. We have seen in this paper that a wide variety of
 materials appear to be in agreement with this theory. No other single theory can explain such
 widely different material classes.  If the theory is correct, realistic calculation of electronic structure looking at the
 right quantities should be predictive as far as whether the material will or will not be a
 superconductor, and give at least a semiquantitative estimate of $T_c$.
 
 \acknowledgements
 The author is grateful to F. Marsiglio for collaboration in much of the work discussed here, as well as to X.Q. Hong and 
 J.J. Hamlin for collaboration in selected portions.

\end{document}